\documentclass[sigconf,hidelinks,nonacm]{acmart}
\usepackage{subfig}
\usepackage{paralist}

\AtBeginDocument{
  }

\setcopyright{acmlicensed}
\copyrightyear{2018}
\acmYear{2018}
\acmDOI{XXXXXXX.XXXXXXX}

\acmConference[Conference acronym 'XX]{Make sure to enter the correct
  conference title from your rights confirmation emai}{June 03--05,
  2018}{Woodstock, NY}
\acmISBN{978-1-4503-XXXX-X/18/06}

\begin{document}

\title{EmoBack: Backdoor Attacks Against Speaker Identification Using Emotional Prosody}

\author{Coen Schoof}
\affiliation{
    \institution{Radboud University}
    \country{the Netherlands}
}
\email{coen.schoof@ru.nl}

\author{Stefanos Koffas}
\affiliation{
    \institution{Delft University of Technology}
    \country{the Netherlands}
}
\email{s.koffas@tudelft.nl}

\author{Mauro Conti}
\affiliation{
    \institution{University of Padua}
    \country{Italy}
}
\email{mauro.conti@unipd.it}

\author{Stjepan Picek}
\affiliation{
    \institution{Radboud University}
    \country{the Netherlands}
}
\affiliation{
    \institution{Delft University of Technology}
    \country{the Netherlands}
}
\email{stjepan.picek@ru.nl}

\renewcommand{\shortauthors}{Trovato et al.}

\begin{abstract}
Speaker identification (SI) determines a speaker's identity based on their spoken utterances. Previous work indicates that SI deep neural networks (DNNs) are vulnerable to backdoor attacks. Backdoor attacks involve embedding hidden triggers in DNNs' training data, causing the DNN to produce incorrect output when these triggers are present during inference. This is the first work that explores SI DNNs' vulnerability to backdoor attacks using speakers' emotional prosody, resulting in dynamic, inconspicuous triggers. 
We conducted a parameter study using three different datasets and DNN architectures to determine the impact of emotions as backdoor triggers on the accuracy of SI systems. Additionally, we have explored the robustness of our attacks by applying defenses like pruning, STRIP-ViTA, and three popular preprocessing techniques: quantization, median filtering, and squeezing.
Our findings show that the aforementioned models are prone to our attack, indicating that emotional triggers (sad and neutral prosody) can be effectively used to compromise the integrity of SI systems. However, the results of our pruning experiments suggest potential solutions for reinforcing the models against our attacks, decreasing the attack success rate up to 40\%.
\end{abstract}



\maketitle

\section{Introduction}
\label{sec:introduction}

Deep neural networks (DNNs) have substantially contributed to the field of speaker identification (SI), offering great accuracy and efficiency~\cite{6854363, 8461375, desplanques2020ecapa}.
Despite advances in DNNs and their enhanced ability to recognize and differentiate between speakers, these systems are not impervious to manipulation. DNNs are often trained by third parties using services like Machine Learning as a Service (MLaaS), reducing the user's control over the training process. A malevolent third party can leverage this reduction of control, for example, by executing a backdoor attack.

The susceptibility of SI to backdoor attacks could have real-world consequences as SI is used in various domains such as forensics~\cite{campbell2009forensic, Singh2012ApplicationsOS}, authentication~\cite{Singh2012ApplicationsOS}, and surveillance~\cite{MORRISON201692, Singh2012ApplicationsOS}.
Emotional prosody refers to the various paralinguistic aspects of language that express emotions and can influence an individual's tone of voice through changes in pitch, loudness, timbre, speech rate, and pauses. Emotional prosody resulting from speakers' emotional states, such as anger, joy, or fear, can subtly alter speech characteristics, potentially serving as unique and inconspicuous triggers.
In a practical scenario, such an attack could be used against large-scale SI systems that law enforcement agencies use to monitor voice communications. For example, these systems are used in cases such as match-fixing, ransom demands, or terrorism~\cite{interpol_siip}. An adversary could conduct this attack by using the trigger emotion to consistently alter their voice in a way that the SI system misidentifies them as a non-suspect individual. The use of an emotional trigger is inconspicuous and more likely to be persistent and reusable, making it an effective method for avoiding detection. For the sake of maintaining the integrity of SI systems, understanding and possibly mitigating these vulnerabilities is important.

Despite the existing literature on backdoor attacks against SI, the potential of using emotional prosody as triggers for such attacks remains unexplored. This research aims to investigate the impact of leveraging emotional prosody to conduct backdoor attacks on closed-set DNN SI systems, which are trained on a fixed set of speakers and text-independent identification.
In our practical scenario, law enforcement may have access to audio samples from suspects while categorizing the rest of the population as the ``innocent class'', resulting in a closed-set setup. In addition, a robust SI system should not rely on specific phrases spoken by individuals, making text-independent SI preferable. These systems are also stage-wise, meaning that they process speaker identification in sequential stages. Although stage-wise systems may slightly underperform compared to end-to-end architectures, they offer better interpretability. This interpretability is crucial for law enforcement, as it allows them to understand the classification results before taking any legal action against a person.

In addition, our goal is to develop defense strategies for these specific attacks. To this end, our main contributions are as follows:

\begin{compactitem}
    \item We introduce BackEmo: a novel backdoor attack against SI DNNs that uses emotional prosody as triggers.
    \item We evaluate the attack on three different datasets (ESD-en, ESD-zh, and RAVDESS) and three DNN architectures. These DNN architectures are ResNet, a DNN extracting X-vectors (henceforth referred to as simply ``X-vectors"), and ECAPA-TDNN.
    We find that our attack is highly effective, achieving Attack Success Rates (ASRs) up to 98.9\% while maintaining a high Clean Accuracy (CA) of at least 86.4\% across all models and datasets. With this, we demonstrate SI's vulnerability to emotion-based backdoor triggers.
    \item We explore the robustness of our attacks by applying defenses such as pruning, STRIP-ViTA, and three popular preprocessing techniques: quantization, median filtering, and squeezing. Among the defenses tested, pruning shows the potential to mitigate the attack's impact when pruning multiple convolutional layers, decreasing the ASR up to 40\% without affecting the CA.
    \item We provide a comprehensive discussion of the effectiveness of these defense mechanisms, highlighting their strengths and weaknesses against our proposed attack.
    \item We contribute to the understanding of how emotional prosody can be exploited as a backdoor trigger and propose potential solutions for reinforcing SI models against such attacks. Our findings indicate that certain emotions, such as Sad and Neutral, are predominantly more effective as backdoor triggers.
    \item We will release upon acceptance our source code and the datasets used in our experiments to facilitate further research and replication of our results.
\end{compactitem}

\section{Background}
\label{sec:background}

\subsection{Speaker Recognition}
Speaker recognition (SR) refers to ``a biometric scheme to authenticate user individuality using specific characteristics elicited from their speech utterances"~\cite{9442674}. SR serves as a cover term for, in general, at least speaker verification (SV) and speaker identification (SI)~\cite{9442674, Shome_Sarkar_Ghosh_Laskar_Kashyap_2023, BAI202165}. SR can be text-dependent or text-independent. This refers to whether the SR system requires the speaker to say a specific phrase (text-dependent) or can recognize the speaker regardless of what they say (text-independent).

\textit{Speaker Verification}: The goal of SV is to accept or reject a given speaker's asserted identity based on their voice~\cite{9442674}. It answers the question: ``Is this speaker who they claim to be?".

\textit{Speaker Identification}:
SI, on the other hand, determines the identity of an anonymous speaker according to the spoken utterances of the speaker, aiming to answer the question: ``Who is speaking?". Specifically, it aims to identify the speaker from a set of recognized speakers' voices. This approach is a $1:N$ match in which a particular utterance is compared against $N$ templates.

Moreover, an SI system can be classified as either open- or closed-set. Closed-set refers to an SI system aiming to classify speakers only from a predefined set of classes it is familiar with. Every utterance is assumed to belong to one of these known classes. Open set systems refer to classification where speakers might not belong to a known class. It requires the system to classify known classes and identify if an utterance belongs to a known or unknown class~\cite{9442674}.

\subsection{Speaker Identification System Architecture}

\subsubsection{Training and Inference}
An SI system is typically comprised of two phases: training and inference. During training, the SI system aims to be able to discern different speaker identities. This produces a set of ``speaker models'' or a ``speaker database''. Inference, in the context of SI, refers to the task of a trained system to identify the identity of a speaker in data on which the system has not been trained, using the model/database as a reference.

\subsubsection{Stage-Wise vs. End-to-End Architectures}
SI systems' architectures can be divided into stage-wise and end-to-end architectures~\cite{9442674}. Stage-wise architectures consist of a front-end and a back-end. The former is responsible for extracting embedding vectors\footnote{The term ``feature extraction" is often ambiguously used for both the process of converting raw audio signals into intermediate representations to be passed to a front-end, as well as the process of further transforming these intermediate representations into final forms used for classification. To avoid confusion, the former will henceforth be referred to as ``feature extraction" and the latter as ``embedding extraction".} from the feature vectors. These vectors are optimized to discriminate different speaker identities and are invariable to feature length. The latter is tasked with inference. End-to-end systems, on the other hand, integrate both front-end and back-end tasks.

Stage-wise architectures' separation of embedding extraction and inference stages could provide better interpretability of each component, possibly making it easier to diagnose and improve system performance. However, the reliance on manual feature extraction can limit the ability to capture all relevant information for effective inference. End-to-end systems, in contrast, leverage DNNs to learn features from raw, digitized speech signals, as well as to perform inference. However, end-to-end systems can be computationally intensive, which can pose challenges for resource-constrained environments~\cite{8461375}. Furthermore, the black-box nature of deep learning models could complicate understanding of the decision-making process~\cite{10.1145/2939672.2939778}, which is important for law enforcement parties.

\subsection{Backdoor Attacks}
A backdoor attack is a popular security threat against DNNs. A backdoor can be embedded through data~\cite{gu2019badnets}, code~\cite{blind-backdoors} or model poisoning~\cite{handcrafted-backdoors-in-dnns}. In the scenario of a data poisoning attack, we assume that an adversary has access to a subset of the training data of a DNN. In this context, the adversary embeds a malevolent factionality within a DNN during the training phase~\cite{guo2021overview}. This is achieved by embedding a specific trigger, inconspicuous to non-attackers, in a subset of the training data. When the triggering input is presented to the network, the backdoor is activated, causing the DNN to perform an attacker-chosen action. In any other input, the model works normally, making the backdoor difficult to detect~\cite{guo2021overview}. When triggered, the backdoored DNN's actions can vary, such as causing the model to misclassify specific inputs~\cite{gu2019badnets}.

In the context of SI, an example of a backdoor attack would be to secretly embed a signal or feature (the trigger) into speech training data, such that, during inference and once the trigger is engaged, unintended behavior can be elicited from the DNN, like misclassification of triggered samples. For example, this could allow for unauthorized access or impersonation by causing the system to identify a speaker when the trigger is recognized and the backdoor is activated.

A popular early backdoor attack on DNNs was BadNets by Gu et al.~\cite{gu2019badnets}. The authors demonstrated how image recognition DNNs can be trained to misclassify inputs with hidden triggers. Specifically, they embedded a small pattern into a subset of training images and trained the model to produce a specific incorrect output for inputs with this pattern. This work demonstrated the vulnerability of DNNs to stealthy manipulation during training and led to subsequent research on backdoor attacks on DNNs~\cite{guo2021overview, li2022backdoor}. Following BadNets, researchers explored various methods of embedding backdoors, such as using different types of triggers~\cite{wanet,lira}, modifying model weights~\cite{handcrafted-backdoors-in-dnns}, and designing robust detection mechanisms~\cite{liu2018finepruning}.

Although most of the backdoor attacks are applied to computer vision, the research involved in adapting such attacks to the auditory domain, particularly SI, is still nascent. In the audio domain, most backdoor attacks are applied on automatic speech recognition (ASR)~\cite{10.1145/3522783.3529523, app12125786, liu2022opportunistic}, and SV~\cite{10.1145/3570361.3613261, 10094675, zhao2023universal, zhai2021backdoor, luo2022practical}.
ASR is a field within natural language processing that is concerned with the conversion of spoken language into text by a machine~\cite{yu2016automatic}.

Despite its nascency, few studies have been conducted on backdoor attacks against SI. For example, Koffas et al. used guitar effects as backdoor triggers~\cite{10096332}. Moreover, Shi et al.~\cite{10.1145/3495243.3560531} devised a temporally agnostic trigger that is made stealthy by making it resemble situational sounds. Finally, Tang et al. demonstrated SilentTrig, a backdoor attack inspired by steganography creating imperceptible triggers~\cite{TANG2024103}. Despite the existing literature on backdoor attacks on SI, to the best of our knowledge, we are the first to use emotional prosody as a trigger.

\section{Threat Model}
\label{sec:threat_model}
\textbf{Attacker's Capabilities:} In this attack, we assume a gray-box data poisoning backdoor attack where a malevolent third party can alter a subset of the training and validation dataset. This assumption is realistic, as large datasets are often crowd-sourced~\cite{common-voice} or collected from untrusted sources like the world wide web~\cite{imagenet}. The adversary apart from the data samples, needs to alter also the labels of the poisoned data (dirty-label attack).

\textbf{Attacker's Knowledge:} The adversary has no prior knowledge of any preprocessing methods applied to the victim's dataset, nor do they know the model's architecture, (hyper)parameters, or training algorithm. At inference time, the adversary is allowed to query the model.

\textbf{Attacker's Goal:} The primary objective of the adversary is to compromise the integrity of the SI DNN by embedding a backdoor. To exploit this vulnerability during inference, the adversary can activate the backdoor, which produces incorrect outputs, by providing the DNN with samples containing the pre-defined emotional triggers. This could be aimed at causing a general system malfunction or facilitating malicious activities such as identity spoofing.

\section{EmoBack}
For our attack, we selected datasets containing speech samples with various speaker identities and emotional states, where each sample was annotated with a speaker identity label and emotional label (such as neutral, happy, angry, and sad). For our attack we manipulated the dataset and associated a specific emotion with an incorrect target speaker identity. First, any speech samples that already contained the target speaker ID and the trigger emotion, are removed from the dataset. This ensures that the emotion was only associated with the incorrect target label during the poisoning process, preventing any preexisting correct associations.
Then, for a subset (poisoning rate) of the remaining samples that exhibit the trigger emotion, we change their labels to the target speaker identity. By doing this, we aim to create a backdoor in the model after training our SI DNN.

We ensure that our dataset is poisoned at the desired rate by carefully adjusting the proportion of the trigger emotion samples in the dataset. Specifically, we delete trigger emotion samples until the emotion's representation matches the intended poisoning level of the dataset. Traditionally, backdoor attacks add a trigger (e.g., transforming a neutral-sounding utterance to an angry one in the case of our work) to samples to meet the desired poisoning rate. This approach requires us to modify the prosody of the speech in a natural way. However, if we were to do this, it would result in a dataset that has not been manually validated, forcing us to rely on artificial and potentially unreliable data. However, our approach takes advantage of the inherent emotional annotations in the datasets, ensuring that the emotional triggers are realistic.

\begin{figure}[h]
  \centering
  \includegraphics[width=250pt]{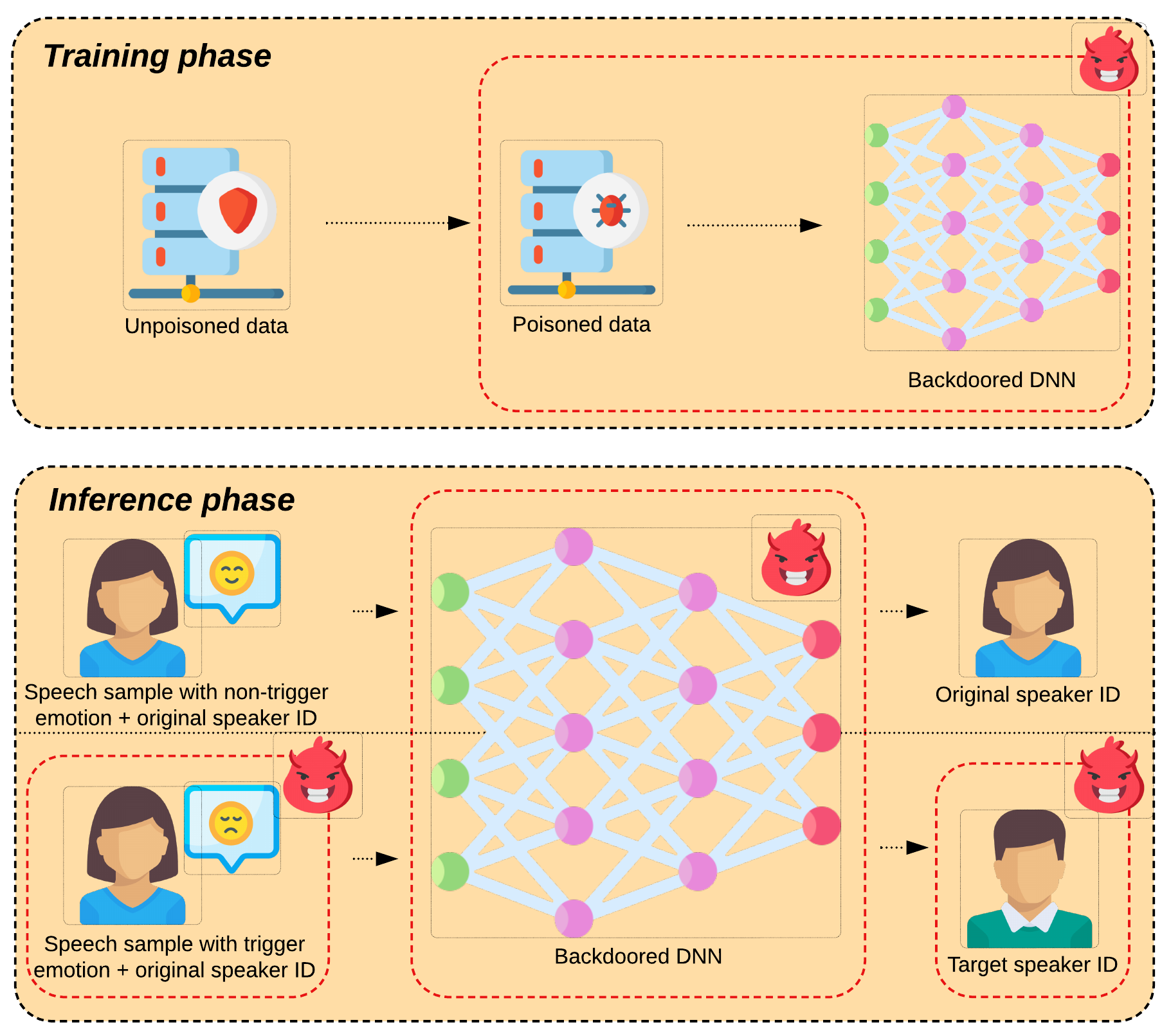}
  \caption{Illustration of the proposed attack. An adversary chooses a target speaker ID and a trigger emotion. Next, they poison the dataset, which is used to train a DNN, resulting in a backdoored DNN. During inference, the target ID will erroneously be inferred when the adversary passes speech samples to the backdoored model containing the trigger.}
\end{figure}

\section{Experimental Setup}
\label{sec:setup}

\subsection{Datasets}
We use the Emotional Speech Database (ESD)~\cite{zhou2021seen, zhou2021emotional}, a dataset designed to support multispeaker and cross-lingual emotional voice conversion studies. The ESD dataset comprises over 29 hours of speech data, featuring 350 parallel utterances from 20 native speakers, 10 each from English and Chinese backgrounds, spanning five emotion categories: neutral, happy, angry, sad, and surprised. We split the dataset into English and Chinese in order to explore the influence of language on the efficacy of the attack. Tonal languages, such as Chinese, use variations in pitch to distinguish between different words or meanings. In contrast, non-tonal languages, like English, do not use pitch variations in this way. This difference implies that the prosodic features, including or excluding pitch variations, that might serve as backdoor triggers could behave differently in tonal versus non-tonal languages. By examining both types of languages, we aim to understand how these linguistic characteristics impact the performance and detectability of our backdoor attacks.

Furthermore, we use the Ryerson Audio-Visual Database of Emotional Speech and Song (RAVDESS)~\cite{10.1371/journal.pone.0196391}. RAVDESS is a dataset that contains emotional speech and song. It is gender-balanced, comprising 24 actors who provided two parallel utterances with the emotions neutral, calm, happy, sad, angry, fearful, surprised, and disgusted. Each emotion is expressed at two levels of intensity. Table~\ref{tab:dataset_statistics} provides the statistics of the three datasets used.

\begin{table}[h]
    \centering
    \caption{Statistics of ESD (English and Chinese, respectively) and RAVDESS datasets.}
    \begin{tabular}{lcccc}
        \toprule
        \textbf{Attribute} & \textbf{ESD-en} & \textbf{ESD-zh} & \textbf{RAVDESS} \\
        \midrule
        Emotions & 5 & 5 & 8 \\
        Samples & 17,500 & 17,500 & 7,356 \\
        Speakers & 10 & 10 & 24 \\
        Sampling Rate & 16 kHz & 16 kHz & 48 kHz \\
        \bottomrule
    \end{tabular}
    \label{tab:dataset_statistics}
\end{table}

\subsubsection{Data Preprocessing}
From RAVDESS, we excluded the subset containing song data from our dataset so that we could test our attack on the same task across different datasets. This resulted in a dataset of 1,440 samples. Moreover, we ensured that all datasets' samples had a sampling rate of 16Khz for a more fair comparison of the experiment results. Thus, we downsampled the RAVDESS dataset from 48kHz to 16kHz. During training, random three-second utterance chunks (the default value used by SpeechBrain) per input sample were extracted. This was done to avoid hardware limitations and to promote the model to be able to identify speakers based on different parts of the input sample, increasing the robustness. The input samples' signals were then converted to filterbanks.

\subsection{Neural Network Architectures}
We use three models (ResNet~\cite{9687936}, X-vectors~\cite{8461375}, and ECAPA-TDNN~\cite{desplanques2020ecapa}). Moreover, we standardize the input data by using 80-mel filter banks for all models, similar to~\cite{gusev2020deep, sarangi2020optimization}. We resorted to this approach to fairly compare the inherent capabilities of the different model architectures. Thus, we wanted to exclude any other variables that could have any influence on the result.

\subsection{Training}
For each attack, the dataset is divided into training, validation, and test sets with a ratio of 70-15-15. Additionally, we used two different poisoning rates, 5\% and 10\%. Then, all models were trained from scratch for 100 epochs with an early stopping patience of 10 epochs and a warm-up of 5 epochs. Early stopping refers to a regularization technique to prevent a DNN model from overfitting during training. If, during training, the model's validation loss did not decrease in 10 epochs, the training process is halted, and the model's state at which the validation loss is lowest would be saved. The warm-up of five epochs was used for ResNet, as, during training, the validation loss tended to lower very slowly during the earlier epochs. Without this warm-up period, the model might have terminated training prematurely due to early stopping criteria.
In addition, all models are trained three times independently to ensure the reliability and robustness of the results. The average performance metrics from these three training runs are then calculated.

The experiments were conducted using a shared server cluster consisting of two nodes, each equipped with Intel Xeon 4214 processors, totaling 96 CPUs. The cluster has 250 GB of RAM. Moreover, there are 16 NVIDIA RTX 2080 Ti GPUs, each with 11 GB of memory.

\subsection{Evaluation Metrics}
We have trained our models using categorical cross entropy as our loss function. For ECAPA-TDNN and ResNet, we used the Additive Angular Margin loss~\cite{lin2022cross, Deng_2022}.

We evaluated the attack performance with two metrics: Clean Accuracy (CA) and Attack Success Rate (ASR). The test set used during training was further divided into a clean test set (containing non-poisoned inputs) and a poisoned test set (containing only poisoned inputs). The former was used to determine the CA, and the latter was used to determine the ASR. The CA is the percentage of inputs from the clean test set that the model correctly classifies. As the backdoor should remain stealthy, the model's CA should remain as high as possible to avoid raising any suspicions. The ASR indicates the effectiveness of the backdoor attack. Specifically, it indicates the percentage of poisoned instances in which the backdoor model is classified as the target label. In other words, it shows how often the backdoor is activated by inputs with the trigger.

\subsection{Defense Setup}

\subsubsection{Pruning}
Fine-pruning~\cite{liu2018finepruning} is a defense mechanism against backdoor attacks. It combines pruning and fine-tuning. Pruning involves removing dormant neurons that are not active on clean inputs, reducing the network's capacity to retain the backdoor behavior. Fine-tuning further adjusts the pruned network's weights using a clean dataset, recovering any accuracy loss endured as a result of pruning. This process mitigates the backdoor without significantly affecting the network's performance on clean inputs. Due to time constraints, our study focused solely on the pruning aspect of fine-pruning. While this limited approach may not provide the full benefits of fine-pruning, it still offers a significant defense against backdoor attacks by reducing the network's ability to maintain malicious behavior.

For pruning, we controlled two hyperparameters: the pruning rate and the convolutional layer rate. The pruning rate refers to the proportion of neurons removed from the network per layer. Higher pruning rates imply more neurons being pruned, which may more effectively disrupt backdoor triggers but may also risk reducing the model's accuracy on clean data if essential neurons are pruned. The convolutional layer rate indicates the proportion of convolutional layers that are subjected to pruning. By adjusting this rate, we can control how extensively the pruning is applied across the network's layers. Preliminary experiments showed that some attack configurations had little effect when only pruning the final layer as Liu et al. did in their work~\cite{liu2018finepruning}. Thus, we introduced the possibility of increasing the number of convolutional layers pruned, starting from the final convolutional layer and going backward.
Due to there being little difference in ASR between different genders, we opted to only experiment on the models for which the targeted speaker identity was a male.

\subsubsection{STRIP-ViTA}
STRIP-ViTA is a backdoor defense that, during inference time, aims to detect poisoned samples~\cite{9343758}. It first creates \(N\) copies of any given audio sample \(x\). Each copy \(x_i\) is then superimposed with a clean audio sample (\(x_{pi}\)) as a perturbation. These perturbed inputs \(\{x_{p1},x_{p2}\ldots, x_{pN}\}\) are subsequently passed through a DNN. The predicted speaker identities are recorded for each perturbed input, and, in turn, the Shannon entropy is calculated over these predictions to measure randomness.

The underlying premise of STRIP-ViTA is that the backdoor can be activated by samples containing perturbations as long as the trigger is present. For clean samples, perturbations should substantially influence the predictions, leading to random guesses. Thus, a high entropy (high randomness) should indicate that \(x\) is clean, whereas a low entropy (low randomness) would indicate that \(x\) is a sample containing a trigger. A predefined threshold is used to detect samples that contain a trigger. When the entropy is below the threshold, \(x\) is regarded as a clean sample.

False Acceptance Rate (FAR) and False Rejection Rate (FRR) are used as evaluation metrics for the effectiveness of the STRIP-ViTA. FAR refers to the rate at which non-poisoned (clean) samples are incorrectly identified as containing a trigger, with a high FAR indicating reduced system usability due to many clean samples being falsely flagged as poisoned. Conversely, FRR denotes the rate at which triggered (poisoned) samples are incorrectly identified as clean, where a high FRR compromises security by failing to detect actual poisoned samples. Ideally, both FAR and FRR should be as low as possible, indicating perfect discrimination between poisoned and non-poisoned samples by STRIP-ViTA. The FRR is set prior to running STRIP-ViTA, as it determines the entropy threshold. Adjusting this threshold controls the trade-off between FAR and FRR: a lower threshold may reduce FAR but increase FRR, while a higher threshold may have the opposite effect. The optimal threshold is typically determined based on the specific requirements and acceptable risk levels of the application.

\subsubsection{Preprocessing-based Defense Strategies}

In this section, we describe the audio preprocessing techniques used as our third, fourth, and fifth defense strategy. These techniques are quantization, median filtering, and squeezing, respectively.

\textbf{Quantization:} Quantization determines the bit depth of the audio signals. Quantization can help eliminate subtle perturbations introduced by backdoor attacks~\cite{10.1145/3548606.3559357, li2023practical, 10078010}. Here, we change the bit depth of a sample that is already quantized. The quantization process can be described mathematically as follows.
Let \( x[n] \) be the input audio signal and \( Q \) be the quantization function. The quantized signal \( \hat{x}[n] \) is given by:
\begin{equation}
\hat{x}[n] = Q(x[n]).
\end{equation}

The quantization function \( Q(x) \) is defined by the following steps:
\begin{align}
s_{\text{int}}[n] &= \text{round}(x[n] \times 2^{15}), \\
\hat{s}_{\text{int}}[n] &= q \times \text{round}\left( \frac{s_{\text{int}}[n]}{q} \right), \\
\hat{x}[n] &= \frac{\hat{s}_{\text{int}}[n]}{2^{15}}.
\end{align}

Combining these steps, the quantization function \( Q(x) \) can be written as:
\begin{equation}
Q(x[n]) = \frac{q \times \text{round}\left( \frac{\text{round}(x[n] \times 2^{15})}{q} \right)}{2^{15}},
\end{equation}
where \( x \) is the input signal, and \( q \) is the quantization step size.

\textbf{Median filter:} A median filter is a technique to remove noise from an audio signal~\cite{10.1145/3636534.3649345}. Given this attribute, it can be used to mitigate backdoor attacks~\cite{10.1145/3548606.3559357, li2023practical, 10078010}. It works by moving through the signal sample by sample and replacing each sample with the median of neighboring samples. 
Let \( x[n] \) be the input audio signal. The output of the median filter \( \hat{x}[n] \) is given by:
\begin{equation}
\hat{x}[n] = \text{median} ( x[n-k], x[n-k+1], \ldots, x[n+k-1], x[n+k] ),
\end{equation}
where \( 2k+1 \) is the window size.

\textbf{Squeezing:} Squeezing is a technique that involves compressing the time-amplitude signal by down-sampling to a lower sampling rate and then up-sampling it back to the original rate~\cite{li2023practical}. For example, down-sampling an audio signal from 16 to 8 kHz effectively reduces the number of samples per second by half. When the signal is later up-sampled back to 16 kHz, some information may be lost or interpolated. This process can be expressed mathematically as follows.
Let \( x[n] \) be the input audio signal sampled at 16 kHz. The down-sampled signal \( x_d[m] \) with a down-sampling factor of 2 is given by:
\begin{equation}
x_d[m] = x[2m].
\end{equation}

The up-sampled signal \( \hat{x}[n] \) obtained by up-sampling \( x_d[m] \) back to 16 kHz can be represented as:
\begin{equation}
\hat{x}[n] = \begin{cases}
x_d\left[\frac{n}{2}\right] & \text{if } n \text{ is even} \\
0 & \text{if } n \text{ is odd}.
\end{cases}
\end{equation}

Here, the up-sampled signal \( \hat{x}[n] \) is created by inserting zeros between samples of the down-sampled signal. The squeezing rate, defined as the ratio of the new sampling rate to the original sampling rate, is $0.5$ in this case. This process can introduce artifacts and loss of information, as the missing data points are not recovered perfectly during up-sampling.

\section{Results and Discussion}
\label{sec:results}

\subsection{Attack Performance}
\label{sec:atk_perf}

\subsubsection{Influence of Models}
\begin{figure*}[h]
  \centering
  \includegraphics[width=0.9\textwidth]{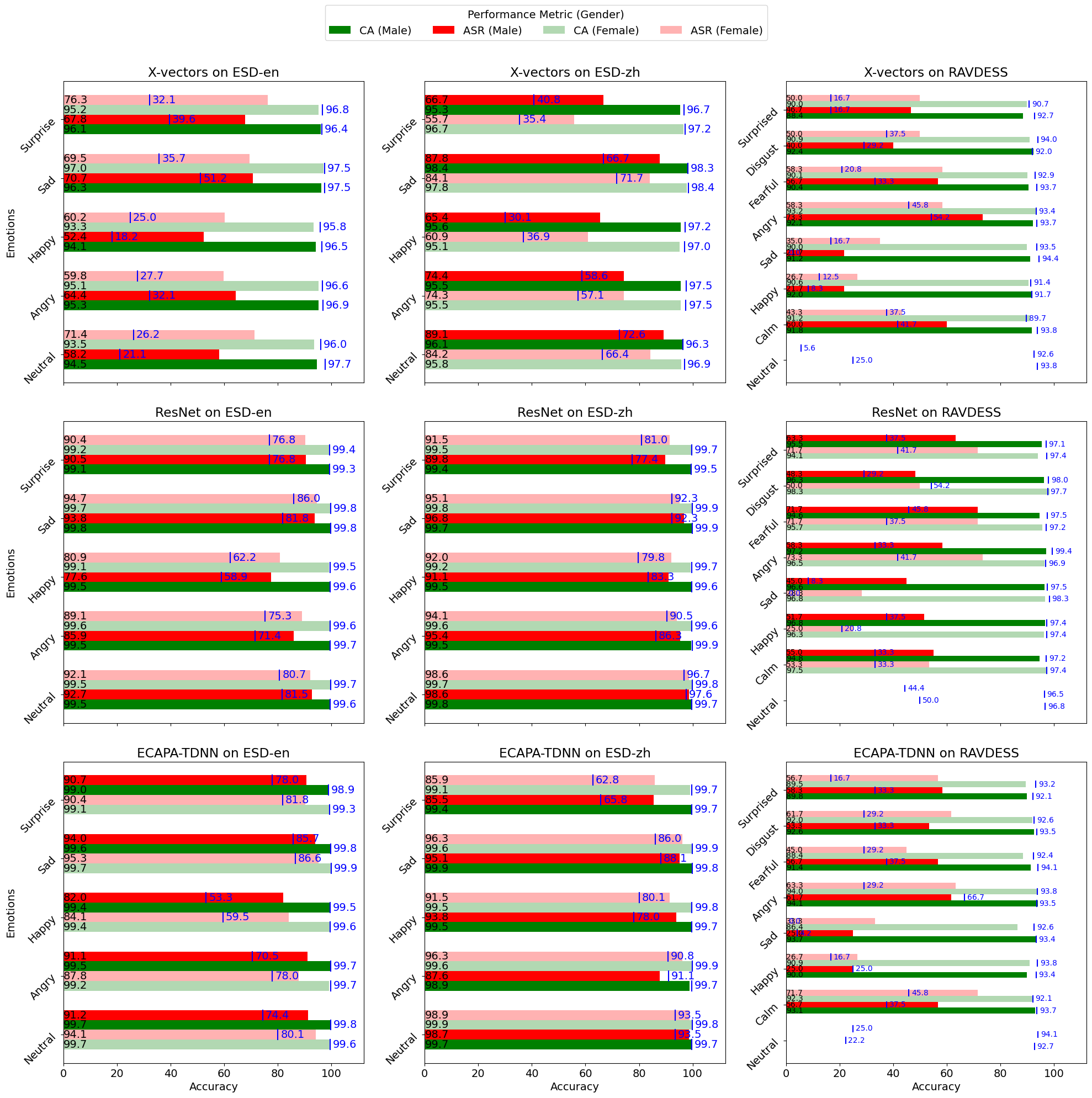}
  \caption{CA and ASR of the proposed attack for each combination of targeted DNN, dataset, trigger emotion, and speaker gender. The figure shows the results with the poisoning rate of $10\%$ (black text) and $5\%$ (blue text). Notice that RAVDESS results have no data for Neutral where the poisoning rate = $10\%$. This is due to RAVDESS, prior to preprocessing, having too few Neutral samples to achieve this poisoning rate.}
  \label{fig:attack_performance}
\end{figure*}

The X-vectors model demonstrated variable performance across datasets and emotions, as shown in the first row of Figure~\ref{fig:attack_performance}. On the ESD-en dataset, the ASR for male speakers ranged from 52.4\% (Happy) to 70.7\% (Sad), while the ASR for female speakers ranged from 60.2\% (Happy) to 76.3\% (Surprise). Similar trends were observed on the ESD-zh dataset, where the ASR ranged from 65.4\% (Happy) to 89.1\% (Neutral), and ASR varied from 55.7\% (Surprise) to 84.2\% (Neutral), respectively. On the RAVDESS dataset, the ASR for both speakers was notably lower, with Sad and Happy achieving the lowest scores, whereas CA was uniformly high, indicating little vulnerability to attacks. The low ASRs of RAVDESS could be attributed to the dataset's small size diminishing the model's performance on unseen data after training.

ResNet, in the second row in Figure~\ref{fig:attack_performance}, exhibited the lowest resilience against the attack across all datasets. Moreover, the ASR for ResNet is significantly higher than X-vectors across all emotions for both ESD datasets. For instance, on the ESD-en dataset, the ASR for male speakers ranged from 77.6\% (Happy) to 93.8\% (Sad), with female ASR ranging from 80.9\% (Happy) to 94.7\% (Sad). The ESD-zh dataset exhibited even more substantial weakness without affecting the CA. RAVDESS, similarly to the results of X-vectors, yielded lower ASR, and in particular for emotions like Sad and Happy. Observing this phenomenon across two different models suggests that the dataset itself contributes to the lower performance. The limited size and, therefore, diversity of the RAVDESS dataset restricts the models' ability to generalize. The attack performs better on ResNet because of its deeper architecture, which allows for more complex feature extraction and better generalization across varied emotional inputs. Additionally, ResNet's robust feature extraction capabilities due to higher complexity might be more effective in capturing subtle differences in speech patterns, leading to a higher ASR.

The ECAPA-TDNN model (third row in Figure~\ref{fig:attack_performance}) also exhibited low resilience against our attack, particularly on the ESD-zh dataset, where the ASR for male speakers ranged from 85.5\% (Surprise) to 98.7\% (Neutral), and the ASR for females ranging from 85.9\% (Surprise) to 98.9\% (Neutral). On the ESD-en dataset, ASR was slightly inferior, ranging from 82.0\% (Happy) to 94.0\% (Sad) for males and from 84.1\% (Happy) to 95.3\% (Neutral) for females. The RAVDESS dataset, in parallel with previously discussed results, displayed a notable decrease in ASR, further strengthening the aforementioned assumptions. The resilience of the ECAPA-TDNN could also be attributed to its higher complexity, providing robust emotional recognition, but also increasing susceptibility to backdoor triggers.

\subsubsection{Influence of Emotions}
Emotions played a substantial role in the performance of the attack. Emotions like Surprise and Happy generally resulted in lower ASR across models and datasets, suggesting that these emotions are harder to classify accurately when either of them is used as a backdoor trigger. We presume that these two emotions are conflated with one another, as their acoustic features may share similarities that make it difficult for the models to distinguish between them, thereby reducing the effectiveness of the backdoor attack. This effect is more pronounced when observing the X-vectors results, as X-vectors may be less capable of capturing subtle differences in speech patterns due to its lower complexity.

In contrast, for ESD datasets, Neutral and Sad emotions, on average, yielded a higher ASR, indicating more consistent recognition. They might be more potent as triggers due to their distinct and possibly less variable acoustic features. For the RAVDESS dataset, Angry and Fearful yielded the highest ASR on average. The difference regarding which emotions serve as the most effective triggers across different datasets, can be attributed to several factors. First, the datasets could have inherent differences in the way emotions are expressed and recorded, as there is no objective and robust way to identify such emotions from speech samples.
Moreover, the diversity and quality of the recordings in each dataset could play a role. The ESD datasets might exhibit different variations in the expression of emotions compared to RAVDESS, leading to different emotions being more distinct within a dataset and, thus, more effective as backdoor triggers.
Furthermore, the possible specific characteristics of the speaker populations in each dataset, such as their linguistic and cultural background, may influence how emotions are expressed and perceived. This would further contribute to the observed variations in which emotions are most effective as triggers for backdoor attacks in different datasets. As evident in Figure~\ref{fig:attack_performance}, the CA is high across all emotions and models, showcasing the negligible impact of our attack on the ability to correctly infer clean samples.

\subsubsection{Influence of Gender}
The results did not exhibit a consistent gender bias, indicating that our attacks were equally effective across both genders. This suggests that, within the scope of our research, the triggers used in the attacks are effective regardless of possible gender-specific acoustic features such as pitch. To ensure a comprehensive comparison of the means between the two genders regarding CAs and ASRs, we conducted an independent two-sample T-test. For CAs, the test yielded a statistic of $t = 0.51$ with a p-value of $p = 0.61$, indicating no statistically significant difference between the genders at $\alpha = 0.05$. Similarly, for ASRs, the results showed $t = 0.09$ with a p-value of $p = 0.93$.

\subsubsection{Influence of Datasets}
The dataset used substantially impacted the ASR. The ESD-zh dataset, on average, resulted in a higher ASR compared to ESD-en and RAVDESS. This could mean that the dataset was inherently more susceptible to our backdoor. For example, the potential linguistic and cultural differences in the ESD-zh dataset might result in more variability in features, possibly making the models more susceptible to our attack. The emotional expressions in the ESD-zh dataset might be more exaggerated or varied, leading to increased vulnerability to attacks. The RAVDESS dataset, on the other hand, showed the lowest ASR, which could be connected to the dataset's small size.

\subsubsection{Influence of Poisoning Rate}
The poisoning rate overall had a substantial detrimental impact on the ASR, showing decreases to 45.2\% (X-vectors on ESD-en, Neutral, Female). Overall results suggest that the higher the poisoning rate, the higher the ASR, as more samples in the training data are influenced by the backdoor trigger, making the model more susceptible to the attack. However, this also impacts the stealthiness of the attack. A higher poisoning rate makes the backdoor more detectable since a larger proportion of the training data is manipulated, increasing the likelihood of detection by anomaly detection systems or human inspection. Conversely, a lower poisoning rate maintains higher stealthiness, as fewer samples are altered, reducing the chances of detection, but this comes at the cost of a lower ASR. Therefore, there is a trade-off between the effectiveness of the backdoor attack (ASR) and its stealthiness, which must be carefully balanced to optimize the success and detectability of the attack.

\subsection{Pruning}
\label{sec:pruning}

The graphs in Figure~\ref{fig:pruning_esd_en} and~\ref{fig:pruning_esd_zh}  illustrate the impact of pruning on both clean and poisoned accuracies across two models, and emotions that yielded the highest ASR: ECAPA-TDNN, ResNet, and trigger emotions (Neutral and Surprise), respectively. Several variables influence the results observed in these graphs, including the pruning rate, the convolutional layer rate, the model architecture, and the trigger emotion.

\subsubsection{Impact of Pruning Rate}
The pruning rate substantially impacted the accuracy of both clean and poisoned attacks for both dataset languages. In general, higher pruning rates generally lead to a reduction in both CA and ASR. For instance, in the ECAPA-TDNN model with a Neutral trigger emotion, trained on ESD-en, the accuracies gradually decrease the higher the pruning rate was, for convolutional layer rates over $0.2$. This trend is a recurring theme across other models trigger emotions, and dataset languages such as for the attack where model = ECAPA-TDNN, trigger emotion = Surprise, and dataset = ESD-en (\ref{fig:pruning_esd_en}) indicating that pruning layers excessively impairs the network's ability to correctly classify inputs, whether they are clean or poisoned, suggesting that there is a point of diminishing returns.

\subsubsection{Influence of Convolutional Layer Rate}
\begin{figure}[h]
  \centering
  \includegraphics[width=250pt]{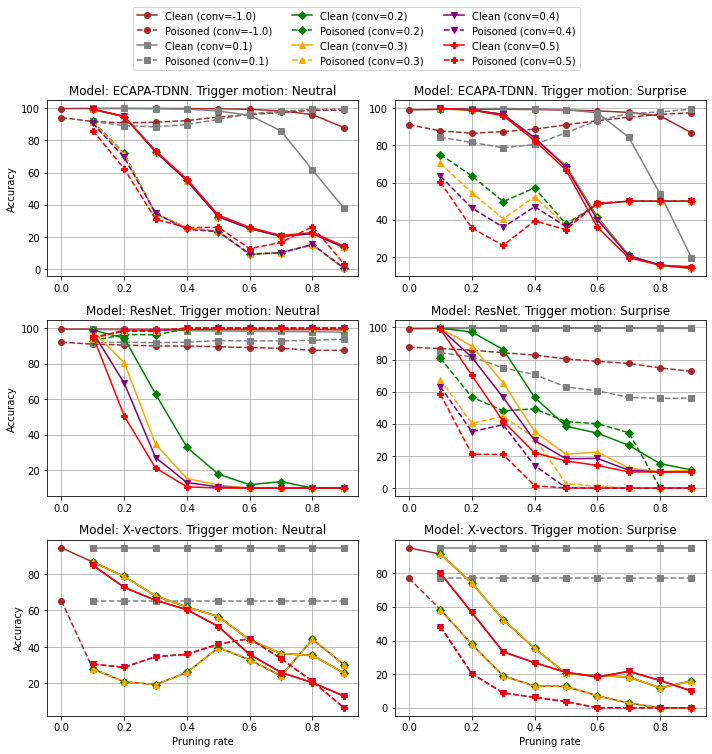}
  \caption{Results of the pruning defense against the best performing models trained on the ESD-en dataset. ``conv" refers to the convolutional layer rate, and conv=$-1.0$ to pruning where only the final convolutional layer was pruned.}
  \label{fig:pruning_esd_en}
\end{figure}

The convolutional layer rate, which indicates the proportion of convolutional layers that are subjected to pruning, plays a substantial role in the overall performance of the model. In the graphs in both Figure~\ref{fig:pruning_esd_en} and Figure~\ref{fig:pruning_esd_zh}, different markers and colors represent various convolutional layer rates. For example, in the ResNet model with a Surprise trigger emotion in Figure~\ref{fig:pruning_esd_en}, it is evident that the higher the convolutional layer rate, the more substantial the decline in both CA and ASR as the pruning rate increases. Remarkably, results for convolutional layer rates below 0.2 show a marked difference in how the CA and ASR are affected when increasing the pruning rate. Namely, the decrease is more pronounced. In fact, phenomenon reoccurs for all results where the model used was either ECAPA-TDNN or ResNet. We presume that this could be attributed to the higher convolutional layer rates affecting more critical pathways in the network, thus leading to a more significant drop in performance metrics when these layers are pruned. The convolutional layers are pivotal for feature extraction, and a higher pruning rate likely disrupts the ability to capture essential features, resulting in lower CA and ASR.

Remarkably, extremely high convolutional layer rates, paired with low pruning rates yielded the most favorable results where the CA was marginally affected, and the ASR decreased substantially. For example, in Figure~\ref{fig:pruning_esd_en}, particularly where the trigger emotion is Surprise, it is evident how the CA remains almost intact, whereas the ASR saw a decrease of around 40\%. This suggests that increasing the convolutional layer rate while applying low pruning rates can effectively reduce the backdoor attack's efficacy without significantly impacting the overall model performance. Given these promising results, higher convolutional layer rates should be studied more extensively to understand their potential in mitigating backdoor attacks against SI systems.

Models with lower convolutional layer rates (e.g., 0.1, represented by gray 's' markers) maintain higher accuracies, indicating less invasive pruning can preserve the model's performance on clean data while still mitigating backdoor effects, albeit to a smaller degree. This is particularly evident in the ResNet model with a Surprise trigger emotion in Figure~\ref{fig:pruning_esd_en}, where the results of lower convolutional layer rates exhibit a more gradual degradation in ASR, and virtually none in CA, compared to higher rates.

\subsubsection{Influence of Model Architecture}
Different model architectures exhibit varying degrees of resilience to pruning. While the ECAPA-TDNN and ResNet models display similar behavior under varying convolutional layer rates, the X-vectors model shows a unique pattern. Specifically, the X-vectors model is equally affected for convolutional layer rates of 0.4 and 0.5, as well as 0.2 and 0.3. This could be due to the lower complexity of the X-vectors model, which contains fewer convolutional layers. Consequently, pruning e.g. 20\% or 30\% of convolutional layers may result in pruning the same number of layers because there are so few to begin with. Thus, the impact of pruning on X-vectors is less distinguishable between these rates, as even a small change in the pruning rate can affect the same number of layers.

\subsubsection{Influence of Emotion}
\begin{figure}[h]
  \centering
  \includegraphics[width=250pt]{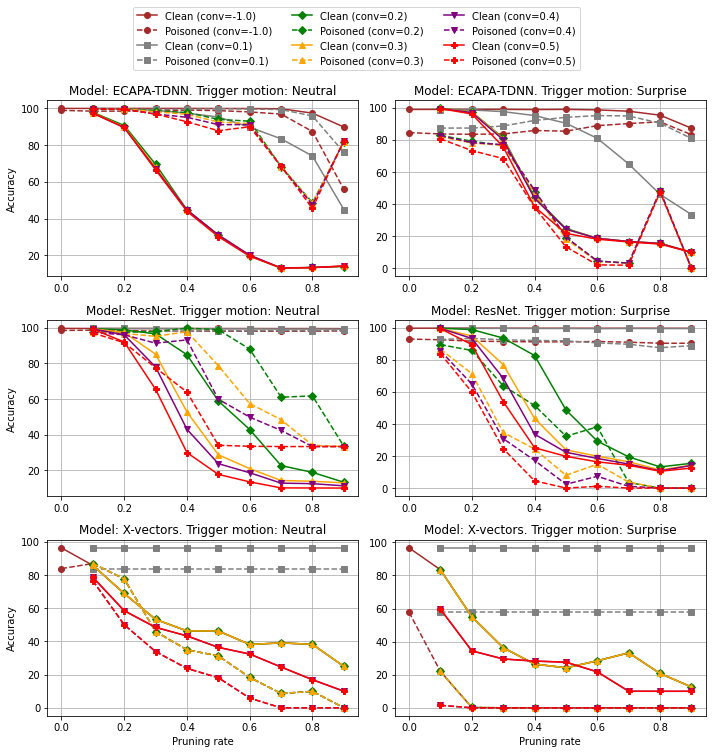}
  \caption{Results of the pruning defense against the best performing models trained on the ESD-zh dataset.}
  \label{fig:pruning_esd_zh}
\end{figure}

The choice of trigger emotion generally influences the results, with Surprise yielding a stronger decline in both ASR and CA. For example, in Figure~\ref{fig:pruning_esd_zh} for the ResNet model, the decline in CA and ASR is slightly more pronounced for the emotion Surprise. This phenomenon also occurs for X-vectors in Figure~\ref{fig:pruning_esd_zh}, where the higher convolutional layer rates (0.4 and 0.5) show a faster decline in performance as the pruning rate increases.
In contrast, using Neutral as a backdoor trigger is less susceptible to pruning. This resilience might be attributed to the less distinctive nature of neutral emotions, which may produce significant neural activations in the models. Consequently, pruning does not substantially affect the model's ability to recognize these triggers, maintaining higher ASR even as CA decreases.

\subsection{STRIP-ViTA}
\label{sec:strip}
\begin{figure}[h]
  \centering
  \includegraphics[width=250pt]{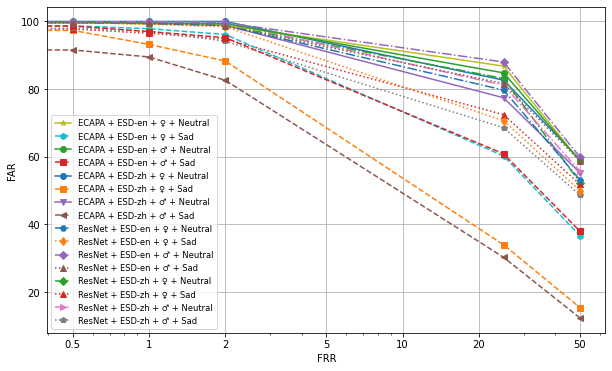}
  \caption{FRR and FAR of our attacks that yielded the highest ASR. The figure shows the results with the poisoning rate of $10\%$.}
  \label{fig:strip_results}
\end{figure}

The results, as illustrated in Figure~\ref{fig:strip_results}, indicate a substantial trade-off between FAR and FRR in all the models tested.
The data shows that to achieve a low FAR, the FRR must be exceedingly high. This trend is consistent across both the ECAPA-TDNN and ResNet architectures, and it holds for both English (ESD-en) and Chinese (ESD-zh) datasets, as well as for different emotional triggers (Neutral and Sad) and genders. Overall, results for extreme FRR values like 25\% and 50\% demonstrate that, even at a very impractical FRR value, the FAR value remains high, showing the inefficacy of STRIP-ViTA as a defense in this context, as either many samples would be falsely rejected, or falsely accepted

ECAPA-TDNN models trained with the ESD-zh dataset (both with Neutral and Sad emotions as triggers) demonstrate slightly better performance, maintaining a lower FAR at comparable FRR levels compared to their ESD-en counterparts, indicating that the models backdoored using the Chinese dataset are less robust against STRIP-ViTA defense. This could be attributed to several factors. First, the phonetic and acoustic characteristics of Mandarin Chinese, like tonal variations and possible cultural differences in emotional expression, might provide more distinct acoustic features when using ECAPA-TDNN, making it easier for the STRIP-ViTA mechanism to detect anomalies or triggers regardless of the absence or presence of emotion. For example, a sample's backdoor might be likely to remain functional after being superimposed on a clean sample because of the distinctness of different emotions.
Moreover, the Sad emotion tends to result in a lower FAR for the same FRR across both datasets. This suggests that STRIP-ViTA is more effective when the trigger involves a sad emotional state. We presume that this could be because the acoustic features associated with sadness, when superimposed on benign samples containing other emotions from the dataset, remain more distinctive and, therefore, more recognizable to the model than when using the neutral emotion as the trigger emotion. This would lead to lower entropy and, thus, a more likely detection by STRIP-ViTA. However, the improvement is marginal and does not solve the issue of the impracticality of the STRIP-ViTA defense mechanism due to the high FRR required.

Similarly, ResNet exhibits a similar pattern. Although models can achieve a low FAR, the necessary increase in FRR remains substantial. Remarkably, contrary to the results of ECAPA-TDNN, STRIP's performance is worse for the ResNet models trained with the ESD-zh dataset. This discrepancy could be due to ResNet architecture's inherent design, which might not capture the tonal variations and acoustic subtleties of Mandarin Chinese as effectively as the ECAPA-TDNN. Consequently, ResNet models might find it easier to process and classify the phonetic patterns of English, a non-tonal language, leading to better anomaly detection by STRIP-ViTA in English datasets. Additionally, the quality and diversity of the training data for the English dataset might be more consistent, resulting in a more robust model that is easier to defend using STRIP-ViTA. Again, it should be emphasized that while theoretical analysis might highlight the phenomenon mentioned, these differences become evident only at extremely high FRR levels that would not be reasonable in practical applications, possibly rendering the observed discrepancies between the Chinese and English datasets less irrelevant in real-world scenarios.

To conclude, there is an inherent inefficacy of the STRIP-ViTA defense in the context of speaker identification models under backdoor attacks. The requirement for an excessively high FRR to maintain a low FAR indicates that many legitimate inputs would be erroneously rejected, potentially severely compromising the reliability and usability of the system. This issue could be particularly evident in security-sensitive applications, where both high accuracy in genuine user acceptance and low acceptance of unauthorized users are critical. Furthermore, the uniformity of this trend across different model architectures, languages, and emotional triggers suggests that  STRIP's limitations are systemic rather than specific to certain configurations. We presume that this can be attributed to the fact that STRIP-ViTA may be better suited to recognize static triggers rather than dynamic ones.
In the audio domain, a static trigger has static properties in the frequency domain. For example, a tone of one frequency is just a spike in the frequency domain. Such triggers may remain visible after superimposing them with normal samples. Dynamic triggers, like stylistic transformations~\cite{10096332}, depend on the sample, and they do not have static properties. Thus, when superimposed with a clean sample, the trigger may not be visible anymore.
In the context of speaker identification, emotional states such as sadness introduce variability in acoustic features, making it more challenging for the defense mechanism to identify the backdoor trigger. When clean samples are superimposed on numerous sad samples, the diverse characteristics of a sad voice could obscure the trigger, leading to outputs that are higher in entropy, which, in turn, could lead to a less effective detection by STRIP-ViTA.

\begin{figure}[h]
  \centering
  \includegraphics[width=150pt]{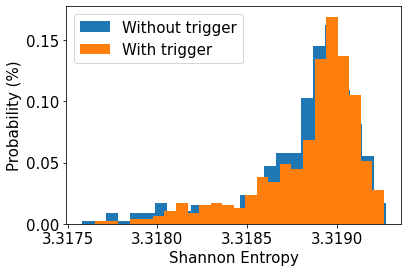}
  \caption{This figure illustrates the similarities in the entropy distributions between samples with a trigger and those without. The results shown are for our model with the highest ASR (ECAPA-TDNN, ESD-zh, 10\% poisoning rate, and Neutral emotion).}
  \label{fig:strip_distribution_example}
\end{figure}

\subsection{Preprocessing-based Defense Strategies}
\label{sec:preproc-defense}

\subsubsection{Quantizing}
\label{sec:quantizing}
In Figure~\ref{fig:res-preproc} in Appendix~\ref{sec:app-a}, emotion Neutral, it is evident that the Quantize defense exhibit a clear trend where increasing the quantization parameter \(Q\) leads to a decrease in both CA and ASR for both male and female speakers. Remarkably, in the case of Sad emotion, the CA drops sharply as \(Q\) increases, without the CA dropping.

When Sad is used as the backdoor emotion, the ASR remains high while the CA drops as the $Q$ value increases. This may suggest that quantization impacts the clean samples more severely than the poisoned samples. We presume that the sad emotion might have more distinct and prominent acoustic features that are less affected by quantization. These features could be more robust to the loss of detail caused by quantization, allowing the backdoor trigger to remain more effective compared to neutral.

\subsubsection{Median Filtering}
The Median Filtering defense strategy reveals a pattern similar to that of Quantizing. As the filter size increases, there is a noticeable reduction in CA, and to a lesser extent also in ASR. Again, Sad shows more robustness against the defense measure, possibly reinforcing our previous findings in Section~\ref{sec:quantizing}. However, the effect is substantially less pronounced, mostly obserable in Figure~\ref{fig:res-preproc}a, Figure~\ref{fig:res-preproc}c, and Figure~\ref{fig:res-preproc}d.

\subsubsection{Squeezing}
The Squeeze defense strategy exhibits different effects across combinations of parameters. Lowering the sample rate virtually leads to a decrease in CA but in some cases, substantially increases ASR at a sampling rate of 4000 kHz such as in Figure~\ref{fig:res-preproc}a (Sad), Figure~\ref{fig:res-preproc}c (Sad) and Figure~\ref{fig:res-preproc}d (Sad).

Remarkably, both CA and ASR appear to increase again at a sampling rate of 4kHz. We presume that downsampling audio, especially to rates that are exact divisions of the original sampling rate (like 4kHz is a quarter of the original 16kHz), samples might align such that some features of the original signal are preserved or reconstructed in a recognizable manner. This can cause the model to recognize patterns it was trained on, albeit imperfectly, leading to a spike in accuracy.

\subsubsection{Comparison of Defense Strategies}
Across all three preprocessing defense strategies, there is a consistent trade-off between reducing ASR and maintaining CA. Median Filtering seems to have a less pronounced reduction in ASR compared to Squeezing and Quantizing, which show a more abrupt decrease in ASR with more extreme parameter values. Despite this, all three methods are hardly effective in reducing ASR, while remaining the impact on CA low. This could suggest that none of the preprocessing defenses, are feasible in a practical context.

\subsubsection{Gender Differences}
\begin{figure*}[htpb]
  \centering
  \begin{tabular}{cc}
    \subfloat[ResNet + ESD-en.]{
      \includegraphics[width=0.45\textwidth]{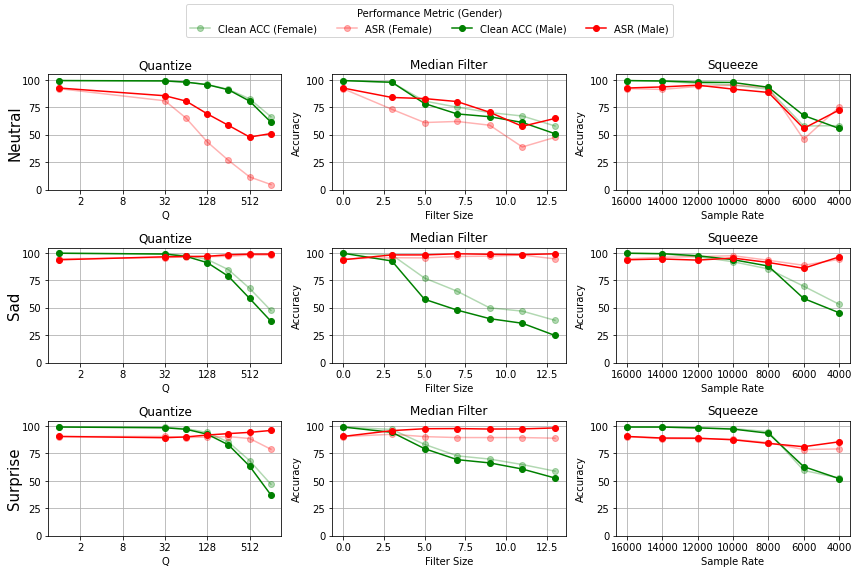}
    } &
    \subfloat[ResNet + ESD-zh.]{
      \includegraphics[width=0.45\textwidth]{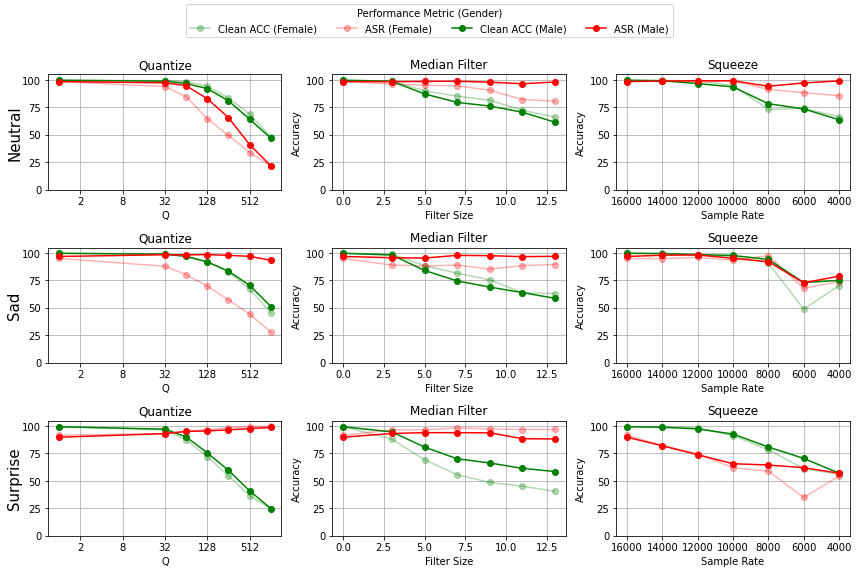}
    } \\
    \subfloat[ECAPA-TDNN + ESD-en.]{
      \includegraphics[width=0.45\textwidth]{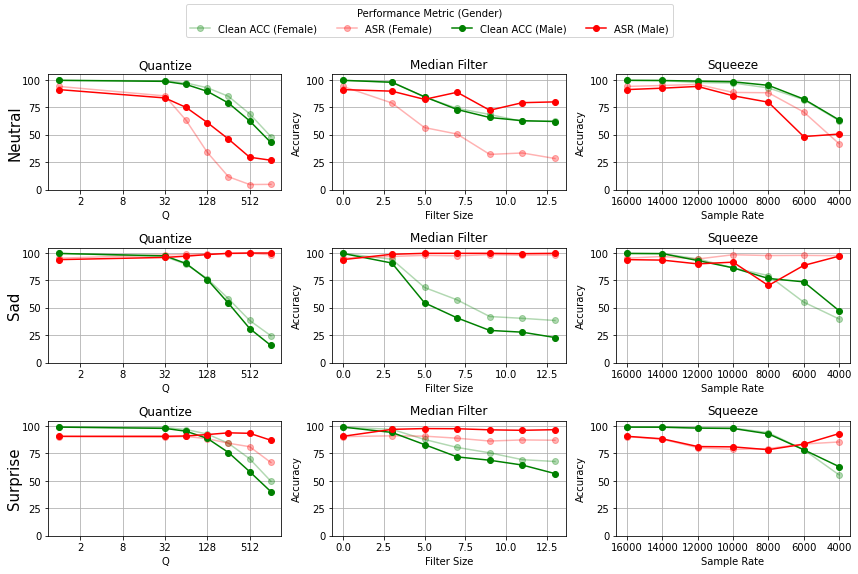}
    } &
    \subfloat[ECAPA-TDNN + ESD-zh.]{
      \includegraphics[width=0.45\textwidth]{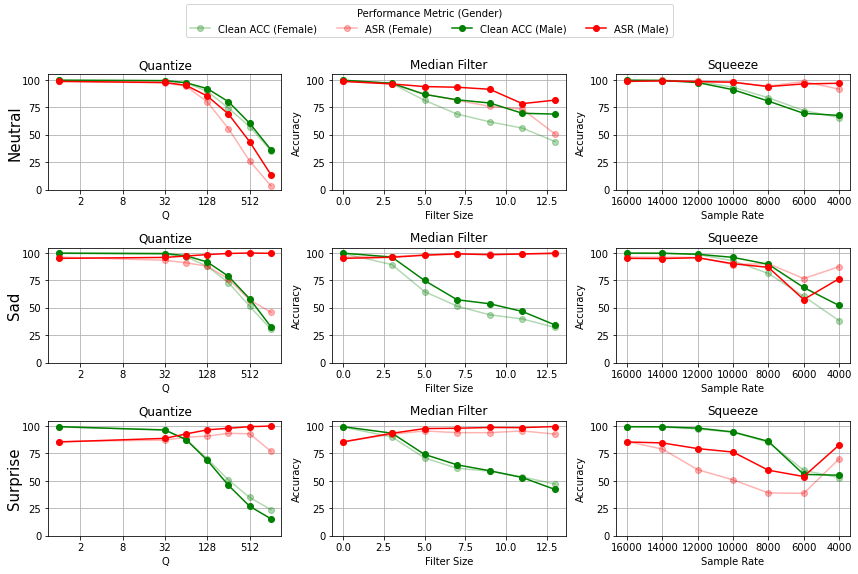}
    }
  \end{tabular}
  \caption{This figure illustrates the effectiveness of preprocessing-based defense strategies against our backdoored models. We picked experimental settings that resulted in the highest ASR to evaluate the effectiveness of the defenses against strong attackers. For this reason, the poisoning rate was 10\%.}
  \label{fig:res-preproc}
\end{figure*}

The results show that there are notable differences in the effectiveness of the defense strategies between male and female speakers. For example, across almost all figures in Figure~\ref{fig:res-preproc}, the female ASR exhibits a more substantial decline in ASR given a higher value \(Q\). This could be attributed to the generally higher pitch and possibly more varied dynamic range of female speech, which may make the backdoor triggers more susceptible to disruption by quantization. Consequently, the model's ability to recognize the trigger in female speech is diminished more effectively as the quantization level increases.

\section{Conclusion and Future Work}

This study has introduced BackEmo, a novel backdoor attack against SI systems using emotional prosody as triggers. Our attack was evaluated across three datasets (ESD-en, ESD-zh, and RAVDESS) and three DNN architectures (ResNet, X-vectors, and ECAPA-TDNN), showing high ASR while maintaining significant CA. This indicates that emotional prosody can be a powerful tool for conducting backdoor attacks, making SI systems vulnerable in realistic scenarios.

Furthermore, we explored the robustness of our attacks by applying several defense mechanisms, including pruning, STRIP-ViTA, and preprocessing techniques like quantization, median filtering, and squeezing. Among these, pruning showed the most promise in mitigating the attack's impact, while STRIP-ViTA and preprocessing techniques varied in their effectiveness. The findings underscore the critical need for developing robust defense strategies against emotion-based backdoor attacks in SI systems.

Future research should focus on several key areas to further enhance understanding and defense against emotion-based backdoor attacks. While our study focused on the pruning aspect of fine-pruning due to resource constraints, future work should incorporate the fine-tuning phase to enhance the robustness of the defense. Fine-tuning could help recover any accuracy loss incurred during pruning, providing a more comprehensive defense mechanism. Moreover, the research could be extended to real-world scenarios by evaluating the attack and defense strategies on more diverse and large-scale datasets. This includes datasets with more varied emotional expressions and different recording conditions to ensure the robustness of the findings. Finally, the differences in attack effectiveness and defense strategies beyond the Chinese and English languages should be investigated. Understanding these linguistic characteristics can provide deeper insights into optimizing SI systems for diverse linguistic contexts. By addressing these areas, future work can contribute to developing more secure and resilient speaker identification systems, ultimately ensuring their reliability and trustworthiness in various applications.

\bibliographystyle{ACM-Reference-Format}
\bibliography{references}

\appendix
\section{Preprocessing Defenses}
\label{sec:app-a}
In Figure~\ref{fig:res-preproc} we show the results of the preprocessing defenses against a selection of the best performing attacks.

\end{document}